# ECG Beats Fast Classification Base on Sparse Dictionaries

Nanyu Li[b], Yujuan Si [a,b*], Di Wang[a], Tong Liu[c], Jinrun Yu[a]

[a] College of Communication Engineering, Jilin University, Changchun, Jilin 130012,China

[b] Zhuhai College of Jilin University, Zhuhai, Guangdong,519041,China

[c] School of Information and Electrical Engineering, Lu dong University, Yantai264025, China

**Abstract**

Feature extraction plays an important role in Electrocardiogram (ECG) Beats classification system. Compared to other popular methods, VQ method performs well in feature extraction from ECG with advantages of dimensionality reduction. In VQ method, a set of dictionaries corresponding to segments of ECG beats is trained, and VQ codes are used to represent each heartbeat. However, in practice, VQ codes optimized by k-means or k-means++ exist large quantization errors, which results in VQ codes for two heartbeats of the same type being very different. So the essential differences between different types of heartbeats cannot be representative well. On the other hand, VQ uses too much data during codebook construction, which limits the speed of dictionary learning. In this paper, we propose a new method to improve the speed and accuracy of VQ method. To reduce the computation of codebook construction, a set of sparse dictionaries corresponding to wave segments of ECG beats is constructed. After initialized, sparse dictionaries are updated efficiently by Feature-sign and Lagrange dual algorithm. Based on those dictionaries, a set of codes can be computed to represent original ECG beats.

Experimental results show that features extracted from ECG by our method are more efficient and separable. The accuracy of our method is higher than other methods with less time consumption of feature extraction

# 1. Introduction

Heartbeat classification is a hot field of Electrocardiogram (ECG) recognition. It's very meaningful to ECG signal records, archiving and diagnosis. Heartbeat classification includes feature extraction and classification. Generally, feature extraction often determines classification results greatly, so it plays an important role in Electrocardiogram (ECG) heartbeats classification system. Features extracted by traditional methods are often from ECG time domain, frequency domain and morphology based on waveform detection point. Unfortunately, those common methods have poor performance, for example: Discrete Cosine transform (DCT), Discrete Wavelet Transform (DWT) and Fast Fourier Transform (FFT) features dimensions (Khorrami H and Moavenian M, 2010) are too high, and also increase time complexity of classification; Morphological and dynamic features (Melgani F and Bazi Y, 2008) are often difficult to be extracted accurately, so these methods cannot perform well especially when the heartbeats data is large. Even though some dimensionality reduction algorithms such as PCA and ICA (Yu S N and Chou K T, 2008) can be adopted to relieve these problems, features extracted by them will lose some important information of heartbeats. Compared to above popular methods, VQ method (Liu T et al, 2016; Liu T et al, 2014) trains a set of dictionaries corresponding to segments of ECG beat and provides a low-dimensionality feature. But the accuracy of ECG classification system based on VQ feature is low. What's more, data used to construct codebook is large during codebook construction and it limits the speed of dictionary learning. To solve these problems, referring to (Wang J et al, 2010.), we learned that sparse codes had better reconstruction performance than VQ codes and were more separable. In this paper, Firstly, we trains a set of sparse dictionaries $[D_1; D_2; \cdots; D_J]$, and each dictionary $D_j$ is corresponding to each segment of training beats. Secondly, we use Feature-sign and Lagrange dual algorithm to update sparse dictionaries, and encode all heartbeats into sparse codes efficiently. Our method can increases the diversity of dictionary structure and makes the feature more efficient. Finally, we use support vector machine (SVM) (Vapnik V N, 1998) as classifier. Compared with neural network (İnan Güler and Übeylı˙ E D, 2005), KNN (Faziludeen S and Sankaran P, 2016), PLSA (Wang J et al, 2013.), SVM can identify all types of heartbeats correctly by a small amount of training data. In our experiment, we achieved a high accuracy. The accuracy was 4% higher than previous methods in MIT-BIH QTDB Database and 3.68% higher than previous methods in MIT-BIH Arrhythmia Database. Furthermore, our method saves more than 50% in time consumption compared with other methods.

By the way, compared to the ECG time domain, frequency domain and morphology feature methods and VQ method, our method is more suitable for mobile medical equipment:

1. Sparse codes are easy to store and take up less memory

2. The results show that higher accurate classification can be obtained with small labeled samples and less execute time by our method, what is more, our method can achieve real-time heartbeat recognition.

In our test, our method can save more than 30 percent in time consumption compared with previous methods

## 2. Organization

The rest of the paper is organized as follows. In Section 3, we introduce the related works: Vector Quantization (VQ) method in heartbeats classification and its disadvantages. In Section 4, in view of the above problems, we introduce our proposed approach: sparse dictionaries learning. In Section 5, we describe how to solve the sparse problem by Feature-sign and Lagrange dual algorithm. By this method, sparse codes of heartbeats can be efficiently obtained. In Section 6, we present ECG Beats classification system based on the proposed approach. In Section 7, we describe MIT-BIH Arrhythmia Database and MIT-BIH QTDB Database used for evaluation of the proposed approach. Experimental results are presented and analyzed. Finally, discussion and conclusion are given in Section.8

## 3. Related Work

Divide ECG beats into a series of segments and learn a set of dictionaries corresponding to them. These dictionaries can be directly utilized in feature extraction for ECG classification. Compared to direct dictionary learning of a whole beat, our method can improve dictionary learning performance. In (Liu T et al, 2016; Liu T et al, 2014), the heartbeats set is $E = [e_1, e_2, \cdots, e_\varphi] \in R^{\Gamma \times \varphi}$, and each heartbeat is divided into J parts by slide windows. For $i^{th}$ heartbeat $e_i = [s_{i,1}, s_{i,2}, \cdots, s_{i,J}]$, based on segments of all heartbeats, a set of K mean dictionaries $B = [D_1, D_2, \cdots, D_J]$ can be constructed. $D_j \in R^{d \times k}$ is the $j^{th}$ dictionary $D_j$ that is corresponding to the $j^{th}$ segment of all the heartbeats $[s_{1,j}, s_{2,j}, \cdots, s_{\varphi,j}]$. Each dictionary includes k clustering centers, the formula is expressed as follows:

$$w_{i,j} = \arg\min_k \left\| s_{i,j} - D_j^{(\kappa)} \right\|_2 \qquad (1)$$

Where $D_j^{(\kappa)}$ represents the $\kappa^{th}$ cluster-center of dictionary $D_j$, and assign code word $w_{i,j} = \kappa$. Eq.(1) can be optimized by K mean or K means++. This process is called Vector Quantization. For each heartbeat $e_i$, It can be represented by a VQ form $a_i = (w_{i,1}, w_{i,2}, \cdots, w_{i,J})$. However VQ process ignores relationships between different cluster-centers, VQ code exists large quantization errors, and the VQ code for heartbeats of the same type might be very different. Therefore essential differences between different types of heartbeats cannot be representative well. When heartbeat dataset is large, we often have no choice but to add more clustering centers to distinguish heartbeat by VQ code words, which leads to increase time complexity of the algorithm. To tackle this problem, we propose to use sparse code instead of VQ code. Consideration can be given to relationship between different clustering centers by sparse code method. As a result, linear combination will be obtained with a small number of clustering centers in the dictionary $D_j$ to represent a segment $s_{i,j}$, and the diversity of the dictionaries structure is increased. We use the dictionaries to encode each beat into sparse code, and sparse codes avoid large quantization errors, which ensure that our codes can be more effective and separable as features of ECG beats. Our method is as shown in Fig.1

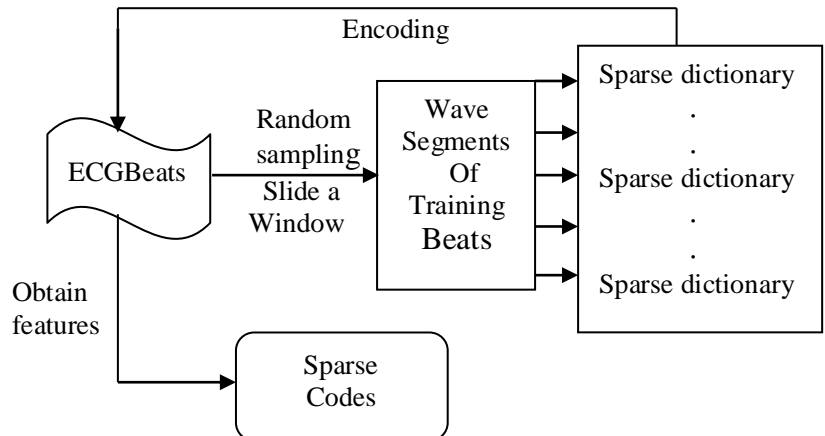

Fig.1- the flowchart of features of ECG beats using the sparse dictonaries

## 4. Sparse dictionaries learning

Sparse dictionary: sparse dictionary learning is a popular algorithm in the image classification, face recognition, and biomedical time series clustering. By sparse dictionary, Sc-SPM (Yang J et al, 2009) achieved 75% accuracy in image classification data set: Caltech 101(Li F F et al 2004), Caltech 256(Griffin G et al 2007); FDDL (Yang M et al 2011)achieved 98% accuracy on Multi-PIE Face database. (Wang J et al, 2013) achieved high accuracy in long term ECG classification. Similar to (Liu T et al, 2016; Liu T et al, 2014), the sparse formula is expressed as follows:

$$\min_{D_j \in R^{d \times k}, W_j \in R^{k \times \varphi}} \sum_{i=1}^{n} \tfrac{1}{2} \left\| s_{i,j} - D_j w_{i,j} \right\|_2^2$$
$$S.t. \quad \forall \kappa = 1, \cdots, k, \left\| d_\kappa \right\|_2^2 = 1 \tag{2}$$
$$and \quad \forall i = 1, \cdots, \varphi, \left\| w_{i,j} \right\|_0 \leq T$$

$w_{i,j}$ is a sparse coefficient, T is Sparseness, and one atom of the Dictionary, $d_\kappa$ has an the square of $l_2$-norm equal to one. For $j^{th}$ dictionary, $W_j = (w_{1,j}, w_{2,j}, \cdots w_{\varphi,j})$ is sparse matrix. KSVD+OMP (Aharon M et al 2006.) can solve Eq. (2), However, Eq.(2) is an NP problem and we can only get the best solution by going through all the solutions. For this reason, KSVD+OMP perform slowly when data is large. Fortunately, in sparse representation, recent theoretical developments (Candes E J and Tao T, 2005; Donoho D L, 2006) reveal that if the solution is sparse enough, solution of Eq.(2) equals to solution of following Eq.(3), which replaces $l_0$ norm with $l_1$ norm, i.e.:

$$D_j = \arg\min_{D_j \in R^{d \times k}, W_j \in R^{k \times \varphi}} \sum_{i=1}^{n} \frac{1}{2} \|s_{i,j} - D_j w_{i,j}\|^2 + \lambda \|w_{i,j}\|_1 \qquad (3)$$

$$S.t. \quad \forall \kappa = 1, \cdots, k, \|d_\kappa\|_2^2 = 1$$

Eq. (3) is a sparse problem. To solve it, we should know how to efficiently construct its sparse dictionary and how to compute its sparse coefficient. What is more, when $W_j$ and $D_j$ are fixed, sparse problem will transform into a convex optimization problem. Thus we use sparse dictionaries $B = [D_1; D_2; \cdots; D_J]$ to encode the heartbeat set, and the diversity of the dictionaries structure is increased. The whole model formula is expressed as:

$$\min_{B \in R^{\Gamma \times k}, A \in R^{k \times \varphi}} \sum_{i=1}^{n} \frac{1}{2} \|v_i - B a_i\|_2^2 + \lambda \|a_i\|_1$$

$$S.t. \quad \forall \kappa = 1, \cdots, k, \|d_\kappa\|_2^2 = 1 \qquad (4)$$

$$B = [D_1; D_2; \cdots; D_J]$$

Because the dictionary B can be updated in Eq. (3), we only need to compute the sparse coefficients *A* in Eq. (4).

## 5. Feature-sign and Lagrange dual algorithm

In this paper, we solve Eq. (3), and Eq. (4) efficiently by Feature-sign and Lagrange dual algorithm (Schölkopf B et al, 2006) in MATLAB platform. The Algorithm 1 summarizes the ECG beats encoding based on sparse dictionaries.

**Algorithm 1: ECG beats encoding by sparse dictionaries**

**Input: heartbeats** $E = [e_1, e_2, \cdots e_\varphi] \in R^{\Gamma \times \varphi}$

**Output: sparse encoding:** $A = [a_1, a_2, \cdots, a_\varphi] \in R^{k \times \varphi}$

**1:** Divide each training beat into J segments,

For each beat, $e_i = [s_{i,1}, s_{i,2}, \cdots, s_{i,J}]$.

**2:** For j=1 to J do

**3:** Solve Eq. (3) by Feature-sign and Lagrange dual algorithm.

**4:** End for

**5:** Update $B = [D_1; D_2; \cdots; D_J]$; Solve Eq. (4) by Feature-sign algorithm.

**6:** Return A

*5.1 Feature-sign Algorithm*

For Eq. (3) and Eq. (4) it's complicated to calculate $W_j$ and A, because $l_1$ norm is non-smooth so that we cannot use Lagrange multiplier method here. Feature-sign search algorithm based on the definition of $l_1$ norm is as follows:

$$\|x_i\|_1 = |x^{(1)}| + |x^{(2)}| + |x^{(3)}| + \ldots \quad (5)$$

Eq. (5) is the $l_1$ norm penalty item, and is not derivable when the element of vector equal to 0. $x^{(\mu)}$ is the $\mu^{th}$ element in vector $x_i$, and we assume we know signs of each element. When only nonzero elements are considered, the problem of the piecewise function with unknown segment points becomes an unconstrained quadratic programming problem:

$$\|x_i\|_1 = x^{(1)} + (-x^{(2)}) + x^{(3)} + 0 + \ldots (if\ x^{(2)} < 0; x^{(1)} > 0; x^{(4)} = 0 \ldots) \quad (6)$$

Refer to Algorithm 2, we can compute $W_j$ and an efficiently. When the $D_j$ and B are fixed, $x_i^{(j)}$ is the j$^{th}$ element of vector $x_i$, considering that $\sigma$ is a vector, so $x_i^{(\sigma)}$ is still a vector.

**Algorithm 2: Feature-sign Algorithm**

**Input:** $E \to Y$ or $s_{i,j} \to y_i$, $Y=[y_1,\cdots,y_n]$; B or $D_j \to D$

**Output:** $X=[x_1,\cdots,x_n]$; $X \to A$ or $W_j$

1. $\mu=\arg\max\limits_{\mu} \dfrac{\partial \|y_i - Dx_i\|_2^2}{\partial x_i^{\mu}}$, $g^{\mu} = \dfrac{\partial \|y_i - Dx_i\|_2^2}{\partial x_i^{\mu}}$

2. If $g^{(\mu)} > \lambda$, then $x_i^{(\mu)} = \dfrac{\lambda - g^{(\mu)}}{D_\mu^T D_\mu}$

   else if $g^{(\mu)} < -\lambda$, then $x_i^{(\mu)} = \dfrac{-\lambda - g^{(\mu)}}{D_\mu^T D_\mu}$;

   else if $|g^{(\mu)}| \leq \lambda$, then $x_i^{(\mu)} = 0$, break;

3. **while true，do**

4. $\sigma = find(x_i \neq 0)$; $D_\sigma = D(:,\sigma)$ ;

   $\dfrac{1}{2}\dfrac{\partial \|y_i^\sigma - D_\sigma x_i^{(\sigma)}\|}{\partial x_i^\sigma} + \lambda sign(x_i^{(\sigma)}) = 0 \to x_{new} = (D_\sigma^T D_\sigma)^{-1}(D_\sigma^T y_i^{(\sigma)} - \lambda sign(x_i^{(\sigma)}))$

5. $sign(x_i^{(\sigma)}) == sign(x_{new})$, then $x_i^\sigma \leftarrow x_{new}$, break

   else if perform a discrete line search

6. **return** $x_i$

*5.2 Lagrange dual Algorithm*

We can use Lagrange dual Algorithm to update dictionaries .when $W_j$ is fixed, updating $D_j$ is a least squares problem with quadratic constraints based on $D_j$, so the Lagrange function is as follows:

$$L(D_j, \lambda) = \|Y - D_j X\|_F^2 + \sum_{\kappa=1}^{k} \lambda^{(\kappa)}(\|d_\kappa\|_2^2 - 1) \tag{7}$$

$\lambda^{(\kappa)}$ ($\lambda^{(\kappa)} \geq 0$) is the element of dual vector $\lambda$; define: $\Lambda = diag(\lambda)$;

$$L(D_j,\lambda) = Tr(Y^TY) - 2Tr(Y^TD_jX) + Tr(Y^TD_j^TD_jY) + Tr(D_j^TD_j\Lambda) - Tr(\Lambda) \tag{8}$$

To obtain the optimal solution $D_\tau$, the first derivative of the Eq. (8) is equal to zero

$$\frac{\partial L}{\partial D_j} = -2YX^T + 2D_jXX^T + 2D_j\Lambda \Rightarrow D_\tau = YX^T(XX^T + \Lambda)^{-1} \tag{9}$$

Substitute Eq. (9) to Eq. (8)

$$L(D_j,\lambda) = Tr(Y^TY) - 2Tr(YX^T(XX^T+\Lambda)^{-1}XY^T) + Tr((XX^T+\Lambda)^{-1}XY^TYX^T) - Tr(\Lambda) \tag{10}$$

So Eq. (10) is a dual optimization problem:

$$\max{}_\lambda \mathbb{R}(\lambda) = \min{}_{D_\tau} L(D_j,\lambda) = Tr(Y^TY - YX^T(XX^T+\Lambda)^{-1}XY^T - \Lambda) \tag{11}$$

We can optimize this dual optimization by Newton's method in Algorithm 3.

**Algorithm 3**: **Lagrange Dual: Maximizing $\mathbb{R}(\lambda)$ by Newton method**

**In put:** $X, Y$;

**Output:** $\lambda_\zeta$

**Initialization:** $\lambda_\zeta$

**1: While $(\sigma < \varepsilon)$ do**

**2: Calculate the gradient and Hessian matrix:**

$$g(\lambda_\zeta) = [\frac{\partial \mathbb{R}(\lambda_\zeta)}{\partial \lambda_\zeta^{(1)}} \cdots \frac{\partial \mathbb{R}(\lambda_\zeta)}{\partial \lambda_\zeta^{(k)}}]^T$$

$$\frac{\partial \mathbb{R}(\lambda_\zeta)}{\partial \lambda_\zeta^{(\kappa)}} = Tr\{[(YX^T(XX+\Lambda)^{-1})^T(YX^T(XX^T+\Lambda)^{-1})]^T \frac{\partial(XX^T+\Lambda)}{\partial \lambda_\zeta^{(\kappa)}}\} - 1 = \|YX^T(XX^T+\Lambda)^{-1}e_\kappa\|^2 - 1$$

Where $e_i = R^k$ is the i-th unit vector, $H(\lambda_\zeta)$ is a nonsingular matrix.

$$H(\lambda_\zeta) = \begin{bmatrix} \dfrac{\partial^2 \mathbb{R}(\lambda_\zeta)}{\partial \lambda_\zeta^1 \partial \lambda_\zeta^1} \cdots \dfrac{\partial^2 \mathbb{R}(\lambda_\zeta)}{\partial \lambda_\zeta^1 \partial \lambda_\zeta^k} \\ \vdots \qquad\qquad \vdots \\ \dfrac{\partial^2 \mathbb{R}(\lambda_\zeta)}{\partial \lambda_\zeta^{(k)} \partial \lambda_\zeta^{(1)}} \cdots \dfrac{\partial^2 \mathbb{R}(\lambda_\zeta)}{\partial \lambda_\zeta^{(k)} \partial \lambda_\zeta^{(k)}} \end{bmatrix}$$

$$\dfrac{\partial^2 \mathbb{R}(\lambda_\zeta)}{\partial \lambda_\zeta^{(\kappa)} \partial \lambda_\zeta^{(\varepsilon)}} = -2Tr[((XX^{\mathrm{T}}+\Lambda)^{-1} e_\kappa e_\kappa^{\mathrm{T}} (XX^{\mathrm{T}}+\Lambda)^{-1} XY^{\mathrm{T}} YX^{\mathrm{T}} (XX^{\mathrm{T}}+\Lambda)^{-1})^{\mathrm{T}} e_\varepsilon e_\varepsilon^{\mathrm{T}}]$$

$$= -2((XX^{\mathrm{T}}+\Lambda)^{-1} XX^{\mathrm{T}} YX^{\mathrm{T}} (XX^{\mathrm{T}}+\Lambda)^{-1})_{\varepsilon,\kappa} ((XX^{\mathrm{T}}+\Lambda)^{-1})_{\kappa,\varepsilon}$$

**3: Do Taylor decomposition;**

$$\mathbb{R}(\lambda) \approx Q(\lambda) = \mathbb{R}(\lambda_\zeta) + g(\lambda_\zeta)^{\mathrm{T}} (\lambda - \lambda_\zeta) H(\lambda_\zeta)(\lambda - \lambda_\zeta);$$

$$\nabla Q(\lambda) = 0, \text{ So } \lambda_{\zeta+1} = \lambda_\zeta - H(\lambda_\zeta)^{-1} g(\lambda_\zeta);$$

$$\sigma = \left\| \dfrac{\lambda_{\zeta+1}}{\lambda_\zeta} \right\|, \quad \zeta = \zeta + 1$$

**4: Return** $\lambda_\zeta$

After maximizing $\mathbb{R}(\lambda)$, we optimize the dictionaries as follows:

$$D_j = YX^{\mathrm{T}}(XX^{\mathrm{T}} + \Lambda)^{-1}, \Lambda = diag(\lambda_\zeta);$$

$$\text{And } B = [D_1; D_2; \cdots; D_J] \qquad (12)$$

By this method, we can update each dictionary effectively, that is because Newton method has second order convergences.

## 6. Our ECG Classification System

ECG beats classification system contains three major parts that are preprocessing, feature extraction, and classification. The process of classification system is shown in Fig.2.

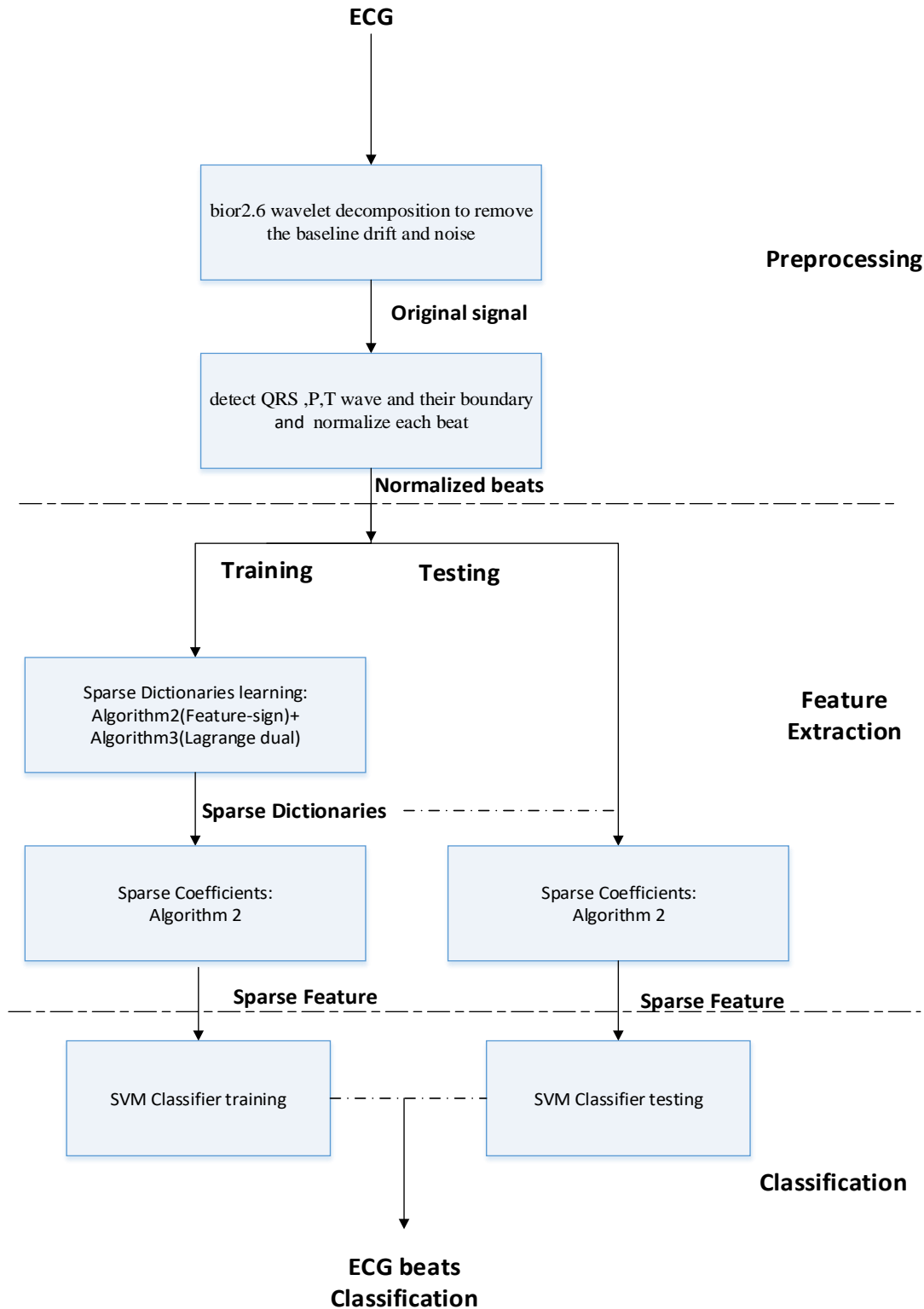

Fig.2 - The flowchart of the ECG classification system

*6.1 Preprocessing*

Based on (Wu D and Bai Z, 2012), we use bior2.6 wavelet decomposition to remove baseline drift and noise, and detect QRS, P, T wave and their boundary. The result of wave detection in ECG is shown in Fig. 3:

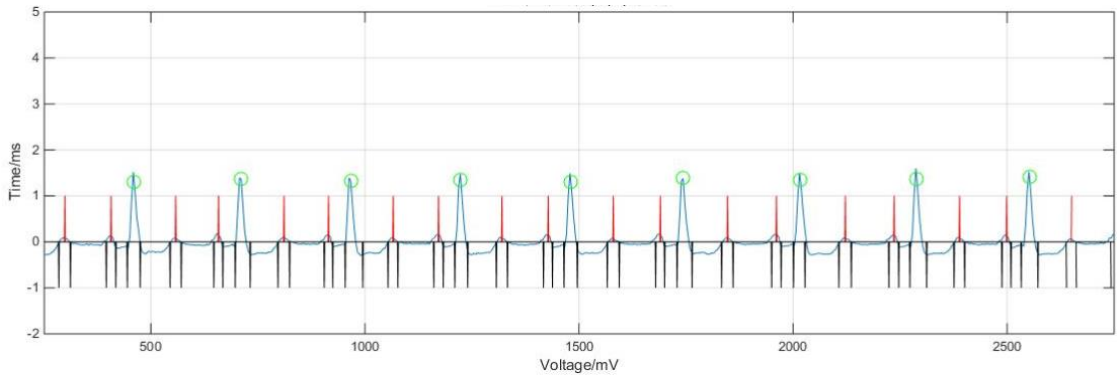

Fig .3-The wave detection in ECG signal

In Figure 3, red line represent peak of P or T wave, green circle represent peak of QRS wave, and black lines indicate boundaries between two waves. We split each heartbeat (the start point of P wave to the end point of T wave), sampling points of which are set to 300, and then normalize them (Wei J J, et al, 2001). We get heartbeat set $E = [e_1, e_2, \cdots, e_\varphi] \in R^{\Gamma \times \varphi}$ .

*6.2 Feature Extraction*

In this part, we can randomly select some beats as training data to learn the dictionary when too many beats segments are extracted from large datasets. This strategy is also employed for the dictionary learning in image and video analysis to reduce the computation of codebook construction (Wei J J et al, 2001). So according to the description in Section 4 and 5, our dictionary learning (Alg.1) is performed to generate the dictionaries $B=[D_1;D_2;\cdots;D_J]$ by training set, and encode the heartbeats as the sparse codes $A=[a_1,a_2,\cdots,a_\varphi]$, this approach improves the diversity of dictionary structures and enables more efficient codes.

*6.3 Classification*

In this part, we use the SVM as the classifier. Compared with neural network (İnan Güler and Übeylı˙ E D, 2005), SVM needs less training data and can achieve higher classification accuracy. The algorithm is based on convex optimization, and can find better optimal solution than non-convex optimization methods such as neural networks. So SVM can improve the performance of heartbeat classification and calculating stability. SVM was proposed by Vapnik et al (Vapnik V N, 1998). It's a supervised learning model based on the idea of the best interval. In our test, we randomly select $z_i \in R^k$ (i=1,2,…,m) from *A* sparse codes set as the training feature.

$$\min_{w,b} \frac{1}{2}\|w\|^2 + C\sum_{i=1}^{m}\xi_i$$
$$S.t. \quad y_i(w^T\Phi(z_i)+b) \geq 1-\xi_i, \forall i \tag{13}$$
$$\xi_i \geq 0, \forall i$$

Where $w^T\Phi(z)+b$ is the defined hyper plane; B is the offset, w is normal vector of the hyper plane, ξ is the relaxation coefficient; C is a penalty factor; we optimize the function to ensure that minimized training errors between two different categories and maximizing geometric intervals $\frac{1}{\|w\|}$; $y_i$ is a label corresponding to a feature $z_i$; Φ is a nonlinear mapping function, the Lagrange function of the SVM optimization is:

$$L_{svm} = \min_{w,b} \frac{1}{2}\|w\|^2 + C\sum_{i=1}^{m}\xi_i - \sum_{i=1}^{m}\beta_i\xi_i - \sum_{i=1}^{m}\alpha_i(y_i(w^T\Phi(z_i)+b)-1+\xi_i) \tag{14}$$

Based on the KKT theorem (Esogbue M.A, 1999), we should have:

$$\frac{\partial L_{svm}}{\partial w} = 0 \Rightarrow w = \sum_{i=1}^{m}\alpha_i y_i \Phi(z_i) \tag{15}$$

$$\frac{\partial L_{svm}}{\partial \xi} = 0 \Rightarrow C = \alpha_i + \beta_i, \forall i \tag{16}$$

$$\frac{\partial L_{svm}}{\partial b} = 0 \Rightarrow \sum_{i=1}^{m}\alpha_i y_i = 0 \tag{17}$$

Substitute Eqs. (15)-(17) into Eq.(14), therefore, to solve Eq.(14) is equivalent to solving a dual optimization:

$$\begin{aligned}L_{svm} = \min_{\alpha} &\frac{1}{2}\sum_{i=1}^{m}\sum_{j=1}^{m}y_i y_j \alpha_i \alpha_j \Phi(z_i)^T \Phi(z_j) - \sum_{i=1}^{m}\alpha_i \\ S.t. \quad &\sum_{i=1}^{m}\alpha_i y_i = 0 \\ &0 \leq \alpha_i \leq C, \forall i\end{aligned} \tag{18}$$

However, the feature mapping $\Phi$ is usually unknown; RBF kernel function can be used instead, which satisfies the Mercer theorem (Vapnik V N, 1998):

$$K(a,b) = \exp(-\gamma \|a-b\|^2) = \Phi(a)^T \Phi(b) \tag{19}$$

where $\gamma$ is kernel parameter that will be automatically set. In this case, the corresponding dual optimization is based on the following formula:

$$\begin{aligned}L_{svm} = \min_{\alpha} &\frac{1}{2}\sum_{i=1}^{m}\sum_{j=1}^{m}y_i y_j \alpha_i \alpha_j K(z_i,z_j) - \sum_{i=1}^{m}\alpha_i \\ S.t. \quad &\sum_{i=1}^{m}\alpha_i y_i = 0 \\ &0 \leq \alpha_i \leq C, \forall i\end{aligned} \tag{20}$$

At last, we only need to find the optimal parameter *C* (penalty coefficient), and $\gamma$ (kernel function coefficient) through PSO (Faziludeen S and Sankaran P, 2016) and then optimize Eq.(20) to obtain hyper plane.

**7. Experiments**

In this section, We do experiments in famous databases: MIT-BIH Arrhythmia Database (Moody G B and Mark R G,2001), MIT-BIH QTDB Database (Goldberger A L et al, 2001), which are widely used in heartbeat classification (Khorrami H and Moavenian M, 2010) (Liu T et al, 2016; Liu T et al, 2014) (Karpagachelvi S et al, 2012). We adopt the same classifier to evaluate the following methods:

•FFT(Fast Fourier Transformation): Former 100 DFT coefficients are extracted as beat feature (Belgacem N et al, 2012) (200dimension for double channel)

•K mean: Referring to (Liu T et al, 2016), K mean dictionaries are built, and each dictionary is corresponding to each wave segment.

•K medioids: Referring to (Liu T et al, 2014), K mean++ dictionaries are built, and each dictionary is corresponding to each wave segment.

•Morphology and Temporal Features(MTF): Referring to (Karpagachelvi S et al 2012), ECG morphology features and three ECG Time-domain features: the QRS complex duration, the RR interval, and the RR interval averaged over the ten last beats are used as features.

•Sparse Dictionaries: According to dictionary learning described in Section 4-5, sparse dictionaries are built, and each dictionary is corresponding to each segment of training beats.

*7.1 MIT-BIH QTDB Database*

7.1.1 Dataset and Experiments settings

The QTDB contains 2-lead ECG signals (ML-II and V5), and both channel signals are adopted in our test, which are available on the Physionet (Laguna P et al, 1997). In our experiment, referring to (Liu T et al, 2016; Liu T et al, 2014), we select the most common 8 categories to classify ECG signal, which are Normal beat (N), Paced beat (/),Atrial premature beat (A), Premature ventricular contraction (V), Fusion of paced and normal beat (f), Fusion of ventricular and normal beat (F),Premature or ectopic supraventricular beat (S), and Right bundle branch block beat (R). However, different categories have different ratios in database. In order to handle this ratio disproportion, some heartbeats are randomly selected from database for training and detailed information is shown in Table1. Except training beats, the rest of database is used for testing.

Table 1 Number of beats for training

| Class | N | / | A | V | f | F | S | R | Total |
|---|---|---|---|---|---|---|---|---|---|
| Training | 350 | 100 | 100 | 200 | 100 | 150 | 100 | 100 | 1200 |

7.1.2 Results and Analysis

As illustrated in Fig.3.Our proposed sparse dictionaries achieves the highest accuracy. Compared with other popular traditional methods, our method outperforms FFT/DWT method by at least 4%, K mean method by at least 9% and K medioids method by at least 5% in accuracy. The accuracy of each category is just as shown in Table.2.

Compared with other algorithm, the accuracy of our proposed method is higher in each category than that of other algorithms. At the same time, they are stable and balanced.

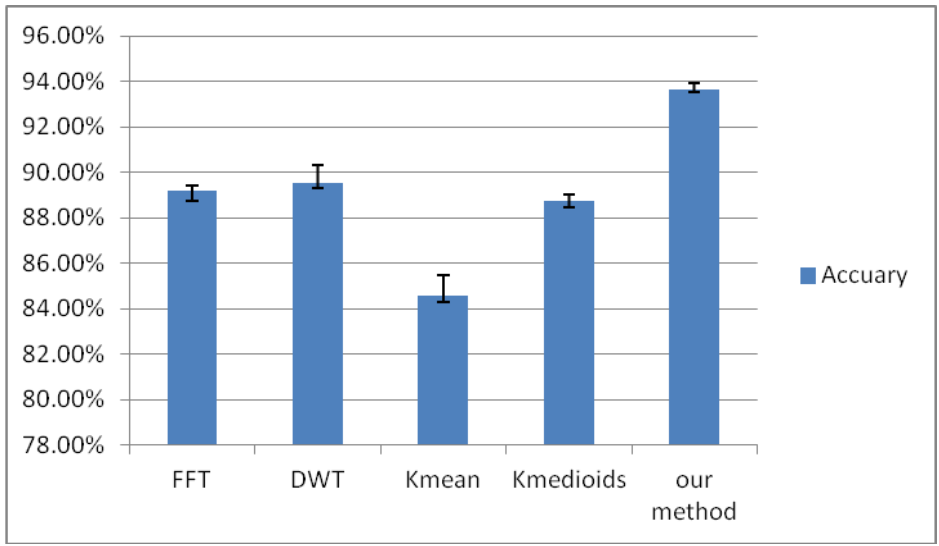

Fig.4- Accuracies achieved on the test beats in MIT-QTDB Database, where error bars represent standard deviations.

Table 2   The accuracies of each category (%) in MIT-BIH QTDB Database

| Method | N | / | A | V | f | F | S | R |
|---|---|---|---|---|---|---|---|---|
| FFT | 95.26 | 87.71 | 97.19 | 91.12 | 93.56 | 81.63 | 82.62 | 84.67 |
| DWT | 94.86 | 88.02 | 92.74 | 88.30 | 96.53 | 81.63 | 79.02 | 95.12 |
| K mean | 95.85 | 85.38 | 92.45 | 82.38 | 93.07 | 73.47 | 63.61 | 90.59 |
| K medioids | 95.92 | 86.62 | 95.43 | 86.43 | 93.07 | 84.69 | 75.90 | 91.99 |
| Our method | 96.01 | 91.91 | 98.42 | 90.90 | 94.55 | 86.73 | 94.43 | 96.17 |

*7.1 MIT-BIH Arrhythmia Database*

7.2.1 Dataset and Experiments settings

MIT-BIH arrhythmia database, as an authoritative database, contains 48 2-lead ECG signals sampled at 360HZ. Similar to (Karpagachelvi S et al, 2012), we select the most common 9 categories, which are: Atrial premature beat (A), Fusion of paced and normal beat (f), Fusion of ventricular and normal beat (F). Left bundle branch block beat (L), Normal beat (N), Paced beat (/), Right bundle branch block beat(R), Premature ventricular contraction (V), Ventricular flutter wave (!). Similar to the QTDB, Some beats are randomly selected as a training set and the detailed information is shown in Table 3. The rest of MIT-BIH is used for testing.

Table 3 Number of beats for training

| class | A | f | F | L | N | / | R | V | ! | total |
|---|---|---|---|---|---|---|---|---|---|---|
| training | 100 | 50 | 50 | 50 | 150 | 50 | 50 | 100 | 50 | 650 |

7.1.2 Results and Analysis

As illustrated in Fig.4, our proposed method achieves the highest accuracy. Compared with other methods, our method outperforms MTF method by at least 4%, K mean method by at least 7% and K medioids method by at least 4% in accuracy. The accuracy of each category is just shown in Table.4. Compared with other algorithm, our algorithm achieves the highest accuracy on each category, and the accuracies are stable and balanced.

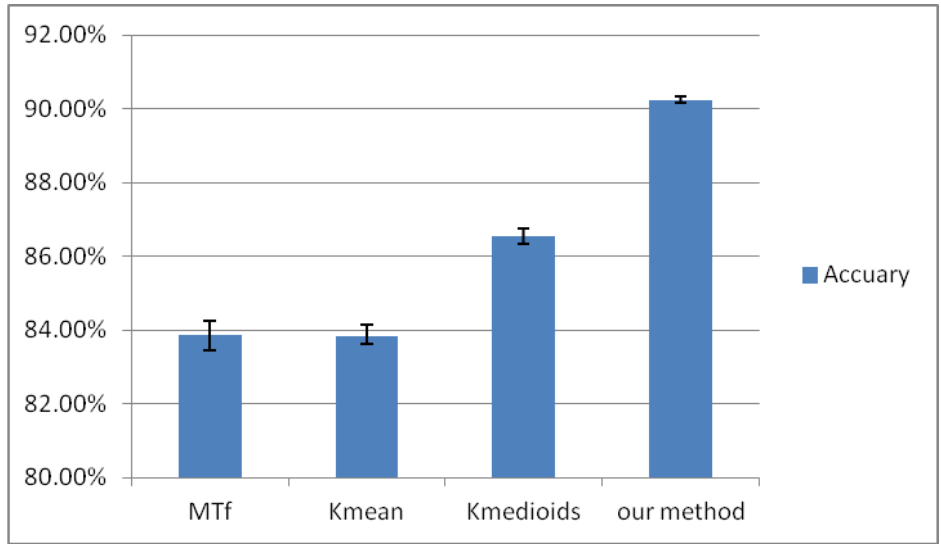

Fig.5- Accuracies achieved on the test beats in MIT-BIH arrhythmia Database, where error bars represent standard deviations.

Table 4 The accuracies of each category (%) in MIT-BIH Arrhythmia Database

| Method | A | f | F | L | N | / | R | V | ! |
|---|---|---|---|---|---|---|---|---|---|
| MTF | 83.19 | 82.11 | 77.12 | 89.94 | 82.19 | 79.11 | 92.03 | 84.48 | 84.60 |
| K mean | 86.22 | 78.56 | 80.61 | 84.77 | 86.61 | 84.29 | 87.09 | 82.12 | 84.20 |
| K medioids | 87.19 | 82.41 | 80.62 | 90.10 | 87.51 | 85.41 | 90.08 | 88.52 | 87.03 |
| Our method | 92.23 | 84.31 | 83.42 | 91.20 | 91.47 | 96.93 | 92.27 | 91.10 | 89.08 |

*7.3 Wilcoxon rank sum test*

Wilcoxon ranksum test is employed to determine whether the accuracy of our proposed method is statistically better than those of other methods. Here accuracies of each method are seen as a group of observations. Results of Wilcoxon ranksum test between each two groups of observations will determine whether to accept null hypothesis. When null hypothesis is accepted, it means that these two groups have the same medians. As shown in Table 5, results of Wilconxon ranksum test between our proposed method and other methods show that it is almost impossible to accept null hypothesis and it means accuracies of proposed method has a different median with those of other methods. Based on these results, we learn that accuracy of our method is not casually higher than those of other methods. So we can conclude that our method statistically outperforms other methods in accuracy.

Table 5 Wilcoxon rank sum test results of each method against sparse dictionaries

| Method | Against to | Probability of accept |
|---|---|---|
| FFT | Sparse dictionaries | $4.6507 \times 10^{-6}$ |
| DWT | Sparse dictionaries | $4.5512 \times 10^{-6}$ |
| MTF | Sparse dictionaries | $4.3507 \times 10^{-6}$ |
| K mean | Sparse dictionaries | $4.3004 \times 10^{-6}$ |
| K medioids | Sparse dictionaries | $4.3207 \times 10^{-6}$ |

*7.4 Algorithm Speed*

In this test, in order to prove the efficiency of our method, we compares the running time (For FFT, DWT, Morphology and Temporal Features, it is the time of the feature extraction; For dictionary learning, it is time of the Dictionary construction and encoding), the tests are conducted on a window7 with Intel Core i5-4200U.1.6GHz and 8GB RAM, and just as shown in Table.6. Our dictionary learning achieves the fast speed, saves more than 50% in time consumption compared with other methods.

Table.6 the time consumption of the feature extraction

| Feature/dictionary | time consumption |
|---|---|
| FFT | 7219.7745s |
| DWT | 3022.2312s |
| Morphology and Temporal Features | 2001.4122s |
| K mean | 1299.5594s |
| K medioids | 899.5594s |
| sparse dictionaries | 412.0002s |

## 8. Conclusion

This paper is aimed at improving the speed and accuracy of the ECG heartbeats classification system. We analyze the drawback of the traditional methods: slow speed and complex calculation. According to our analysis, reasons of these drawbacks include difficultly accurate extraction of morphology, time and frequency features from heartbeats and too many high-dimension features used for training. Although VQ method can solve problems above, its accuracy does not satisfy our requirement. What's more, during training stage, VQ method must use all heartbeats to do unsupervised clustering. As a result the speed of dictionary learning is greatly limited. We propose a method that builds sparse dictionaries corresponding to each segment of training beats and encode all the ECG beats. The advantages of spare coding are dimensionality reduction and accuracy increase, and we don't need large training data. Unlike VQ dictionaries, sparse code has better reconstruction performance than VQ code and is more linearly separable.

The experiment results show that the proposed method has the high accuracy and is capable of reducing the computational complexity of ECG beats classification system and improves the performance of our classifier.

## Acknowledgements

This work was supported by the Key Scientific and Technological Research Project of Jilin Province under Grant [number 20150204039GX and 20170414017GH]; the Natural Science Foundation of Guangdong Province under Grant [number 2016A030313658]; Innovation and Strengthening School Project (provincial key platform and major scientific research project) supported by Guangdong Government [number 2015KTSCX175]; the Premier-Discipline Enhancement Scheme Supported by Zhuhai Government[number 2015YXXK02] and Premier Key-Discipline Enhancement Scheme Supported by Guangdong Government Funds[number 2016GDYSZDXK036]; Natural Science Foundation of China under Grant [number 61702249], and the Natural Science Foundation of Shang Dong Province under Grant [numberZR2016FB18]

**References**

Yu S N, Chou K T, 2008. Integration of independent component analysis and neural networks for ECG beat classification[J]. Expert Systems with Applications, 34(4):2841-2846.

Khorrami H, Moavenian M, 2010. A comparative study of DWT, CWT and DCT transformations in ECG arrhythmias classification[J]. Expert Systems with Applications, 37(8):5751-5757.

Melgani F, Bazi Y, 2008. Classification of Electrocardiogram Signals With Support Vector Machines and Particle Swarm Optimization[J]. Information Technology in Biomedicine IEEE Transactions on, 12(5):667-77.

Liu T, Si Y, Wen D, et al, 2016. Dictionary learning for VQ feature extraction in ECG beats classification[J]. Expert Systems with Applications, 53: 129-137.

Liu T, Si Y, Wen D, et al, 2014. Vector Quantization for ECG Beats Classification[C]//Computational Science and Engineering (CSE), 2014 IEEE 17th International Conference on. IEEE, 13-20.

Wang J, Yang J, Yu K, et al, 2010. Locality-constrained Linear Coding for image classification[J]. 119(5):3360-3367.

İnan Güler, Übeylı̇ E D, 2005. ECG beat classifier designed by combined neural network model[J]. Pattern Recognition, 38(2):199-208.

Faziludeen S, Sankaran P, 2016. ECG Beat Classification Using Evidential K -Nearest Neighbours[J]. Procedia Computer Science, 89:499-505.

Wang J, Liu P, FH She M, et al, 2013. Biomedical time series clustering based on non-negative sparse coding and probabilistic topic model[J]. Computer methods and programs in biomedicine, 111(3): 629-641.

Yang J, Yu K, Gong Y, et al, 2009. Linear spatial pyramid matching using sparse coding for image classification[C] IEEE Computer Vision and Pattern Recognition, 1794-1801.

Li F F, Fergus R, Perona P, 2004. Learning Generative Visual Models from Few Training Examples: An Incremental Bayesian Approach Tested on 101 Object Categories[C] IEEE Computer Vision and Pattern Recognition Workshop, 178-178.

Griffin G, Holub A, Perona P, 2007. Caltech-256 Object Category Dataset[J]. California Institute of Technology.

Yang M, Zhang D, Feng X, et al, 2011. Fisher Discrimination Dictionary Learning for sparse representation[J]. Proceedings, 24(4):543-550.

Aharon M, Elad M, Bruckstein, 2006. A. K-SVD: An algorithm for designing overcomplete dictionaries for sparse representation[J]. IEEE Transactions on Signal Processing, 54(11):4311-4322.

Candes E J, Tao T, 2005. Decoding by Linear Programming[J]. IEEE Transactions on Information Theory, 51(12):4203-4215.

Donoho D L, 2006. Compressed Sensing[J]. IEEE Transactions on Information Theory, 52(4):1289-1306.

Schölkopf B, Platt J, Hofmann T, 2006. Efficient sparse coding algorithms[C] International Conference on Neural Information Processing Systems. MIT Press, 801-808.

Wu D, Bai Z, 2012. An improved method for ECG signal feature point detection based on wavelet transform[C] Industrial Electronics and Applications.1836-1841.

Wei J J, Chang C J, Chou N K, et al, 2001. ECG data compression using truncated singular value decomposition.[J]. IEEE Transactions on Information Technology in Biomedicine A Publication of the IEEE Engineering in Medicine & Biology Society, 5(4):290-9.

J.C. Niebles, H. Wang, 2003. L.Fei-Fei, Unsupervised learning of human action categories using spatial–temporal words, International Journalof Computer Vision 79 (3) 299–318.

Vapnik V N, 1998. Statistical Learning Theory[J]. Encyclopedia of the Sciences of Learning, 41(4):3185-3185.

Esogbue M.A. 1999. Invexity and the KuhnTucker Theorem[J]. Journal of Mathematical Analysis and Applications, 236(2):594-604.

Moody G B, Mark R G,2001. The impact of the MIT-BIH arrhythmia database.[J]. IEEE Engineering in Medicine & Biology Magazine the Quarterly Magazine of the Engineering in Medicine & Biology Society, 20(3):45.

Goldberger A L, Amaral L A N, Glass L, et al, 2001. PhysioBank, PhysioToolkit, and PhysioNet[J]. Circulation, 101(23):215-20

Karpagachelvi S, Arthanari M, Sivakumar M, 2012. Classification of electrocardiogram signals with support vector machines and extreme learning machine[J]. Neural Computing & Applications, 21(6):1331-1339.

Belgacem N, Fournier R, Reguig F B, et al, 2012. Human Authentication System Based on ECG Signal Using FFT and Random Forests[C] International Conference on Informatics & Applications.

Laguna P, Mark R G, Goldberg A, et al, 1997. A database for evaluation of algorithms for measurement of QT and other waveform intervals in the ECG[C] Computers in Cardiology. IEEE, 673-676.